# Weather Warriors:
# A Machine Learning Framework for Climate-Resilient Insurance and Real Estate Decisions

Lang Qin [1], Yuejin Xie [2], Daili Hua [2], and Xuhui Meng [1]

[1] School of Mathematics and Statistics, Huazhong University of Science and Technology

[2] School of Electronic Information and Communications, Huazhong University of Science and Technology

## ABSTRACT

Extreme weather events increasingly threaten the insurance and real estate industries, creating conflicts between profitability and homeowner burdens. To address this, we propose the SSC-Insurance Model, which integrates SMOTE, SVM, and C-D-C algorithms to evaluate weather impacts on policies and investments. Our model achieves 88.3% accuracy in Zhejiang and 79.6% in Ireland, identifying a critical threshold (43% weather increase) for insurance viability. Additionally, we develop the TOA-Preservation Model using TOPSIS-ORM and AHP to prioritize building protection, with cultural value scoring highest (weight: 0.3383). Case studies on Nanxun Ancient Town show a 65.32% insurability probability and a protection score of 0.512. This work provides actionable tools for insurers, developers, and policymakers to manage climate risks sustainably.

# 1 Introduction

## 1.1 Problem Background

Every year, heatwaves, floods, and wildfires bring death and devastation worldwide. These phenomena highlight the profound impact of extreme weather on society, including humans, nature, and even property and insurance owners. In 2020, global insurance losses from natural disasters reached 81 billion, while losses from manmade disasters amounted to 8 billion [1]. The crisis between insurance companies and property owners is becoming evident.

High premiums can lead to a decrease in the number of policies, while low premiums can jeopardize the survival of the insurance company. Therefore, it is particularly important for insurance companies to comprehensively evaluate whether a project is worth insuring. The real estate industry, including property owners and developers, is also affected by insurance companies, which requires them to make adjustments based on the current situation to cope with ongoing developments.

What is more, during the process of changing insurance landscape, certain regions where valuable and significant properties are located may face the risk of being uninsured. Therefore, it is crucial to identify properties that have insurable value in order to provide better protection for them.

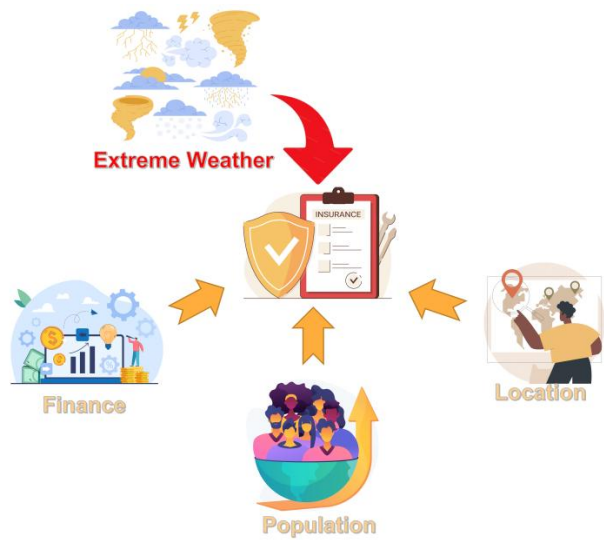

Figure 1: What Matters for Insurance?

## 1.2 Restatement of Problems

In order to alleviate the tense relationship between insurance companies and the real estate industry, it is necessary to consider when insurance companies should be liable for the premium, how property owners can influence this decision, how real estate developers should invest, and how valuable buildings can be protected. Therefore, our team will undertake the following tasks:

- Establish a model to determine the conditions, under which insurance companies can provide coverage for policies, explain when and how assuming risks is worthy, and describe how property owners can influence the decisions.
- Adjust the aforementioned model and utilize the modified model to evaluate where and how to develop real estate in certain location.
- Certain properties within communities will be at risk of being damaged if they are not insured. The establishment of a protection model can assess these properties and propose measures to better protect them.
- Using the established model to analyze a specific historical landmark and propose relevant protection suggestions.

## 1.3 Our Work

In order to avoid complicated description, intuitively reflect our work process, the flow chart is shown in Figure 2:

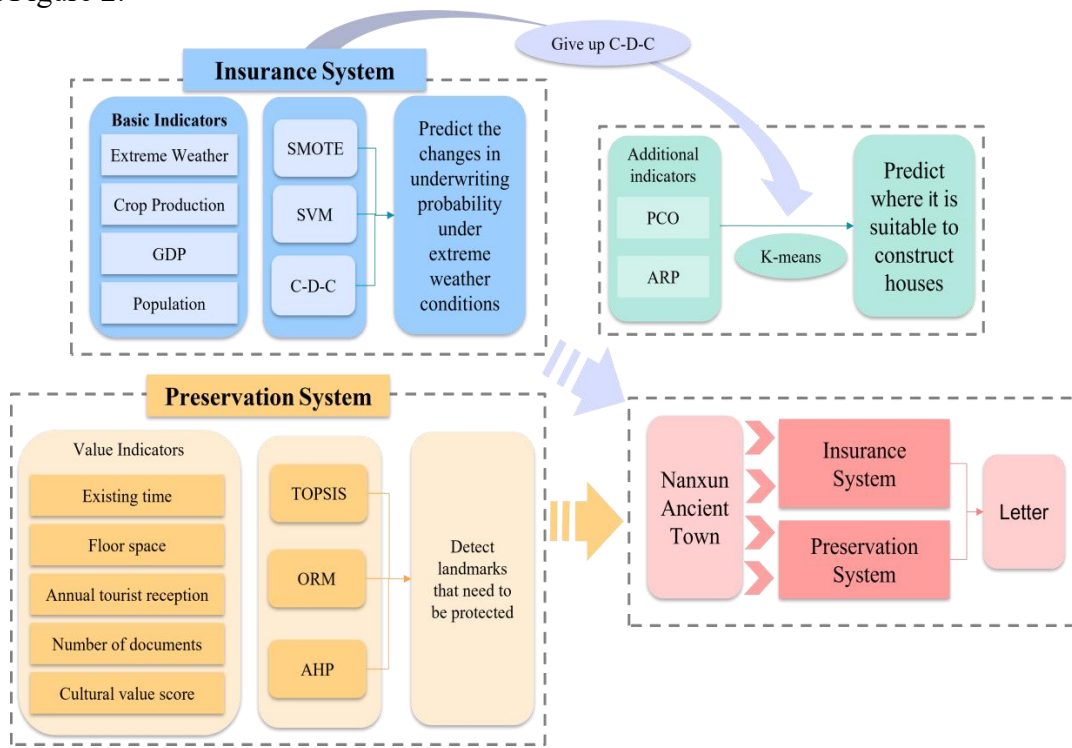

Figure 2: Our Work

## 2 Assumptions and Notations

### 2.1 Assumptions

To simplify our model, we make some assumptions in our paper. The details are as below:

- Assumption 1: Statistics we collect from the website are actual and authoritative.

Justification 1: Our data is from national and global platforms. We believe that the data from these authoritative platforms have undergone rigorous statistical analysis and screening, ensuring their accuracy and reliability.

- Assumption 2: A region under study is a whole unit, regardless of its internal regional differences.

Justification 2: The problem we are discussing is about multiple indicators at the macro level. For the region we have chosen, we believe that the indicators we select can reflect the overall situation of the region as a whole, without being affected by internal differences.

- Assumption 3: The selected indicators in our model is continuous, not experiencing mutations. And we assume the mutual relationships among our indicators are relatively stable.

Justification 3: Our evaluation and prediction models are based on mathematical analysis of collected data. To ensure the data usefulness and model reliability, this assumption is necessary and reasonable.

## 2.2 Notations

Table 1: The Notations in Our Paper

| Symbol | Definition |
| :---: | :---: |
| $n$ | the number of years |
| $m$ | the number of indicators |
| $x_{ij}$ | the $i$th year, $j$th indicator's value |
| $x'_{ij}$ | the normalized value of $x_{ij}'$ |
| $\beta_j$ | Extreme weather's impact on the $j$th indicator |
| $x_{pred_j}$ | the prediction value of $j$th indicator |
| $x_{\max(j)}$ | the maximum of the $j$th indicator across all the years |
| $x_{\min(j)}$ | the minimum of the $j$th indicator across all the years |
| $D_{ij}$ | the distance between indicators of the $i$th year for the $j$th indicator |
| $S_j$ | the importance of each indicator |
| $r_j$ | is the rate of descending sort of important coefficient of $S_j$ |
| $w_j$ | the weights of the $j$th indicator by TOPSIS-ORM |
| $a_j$ | the weights of the $j$th indicator by AHP |
| $Score_j$ | the score of $j$th landmark |

# 3 Insurance Model: SSC Model

In this section, we developed our insurance model with two components. The first component includes the SMOTE algorithm and SVM model. To address the severe imbalance between positive and negative samples in the training set, we employed the SMOTE algorithm to generate negative samples. Due to limited samples, we opted for the relatively simpler SVM to evaluate the positive and negative samples. The second component is the prediction model. By optimizing the C-D model to the C-D-C model, we calculated the correlation between extreme weather and other indicators. Then we can calculate how the other indicators' values change when extreme weather events increase. By incorporating these values into the decision model, we made the final assessment.

## 3.1 Main Factors and Data Selection

Property insurance, as a contractual economic relationship based on property, is one of the important pillars of financial and social security systems. We believe that the problem focusing on how to correctly evaluate and make decisions when the insurance and real estate industries face profitability and burden crises respectively is a comprehensive problem covering multiple fields. Therefore, we selected Extreme Weather, Population [4] [6], Per Capita GDP, Crop Production, and Net Premium Margin as our indicators for modeling and computational analysis. And we consider all these indicators as positive indicators.

3.1.1 Extreme Weather

An Extreme Weather event is rare at a particular place and time of year, with unusual characteristics in terms of magnitude, location, timing, or extent, including heatwaves, cold waves, heavy precipitation, drought, tornadoes, and others [5]. We select relevant records of extreme weather events from the Emergency Events Database (EM -DAT) to address our problem.

3.1.2 Population

Population is one of the crucial geographical factors in a region. Changes in population can cause fluctuations in the demand for insurance in the area, and can also be a factor influencing the structural changes in the insurance market of that region, thereby affecting the corresponding changes in premiums generated in the area.

3.1.3 Per Capita GDP

Per Capita GDP is a macroeconomic indicator that measures the economic development and living standards of a region. Per capita GDP is usually associated with purchasing power, which means that people's purchasing desires will change with the economic environment, thereby affecting the insurance premiums of a region.

3.1.4 Crop Production

Crop Production is an important indicator for measuring the agricultural level of a region. The crop production in a region is correlated and interacts with the local economic development level and population. Additionally, crop production may also affect the insurance structure and insurance policies in the area. Therefore, we choose to include crop production in the evaluation model.

### 3.1.5 Net Production Margin

Given the continuously improving economic level of each region and the long-term growth trend of insurance business volume and insurance profits, we define a metric

$$NPM = \frac{Total\ premium - Claim\ payout}{Total\ premium} \quad (1)$$

It converts the absolute profit and loss situation of the insurance industry into relative profit and loss situation to eliminate the continuous increase of absolute amount caused by the overall increase of economic volume.

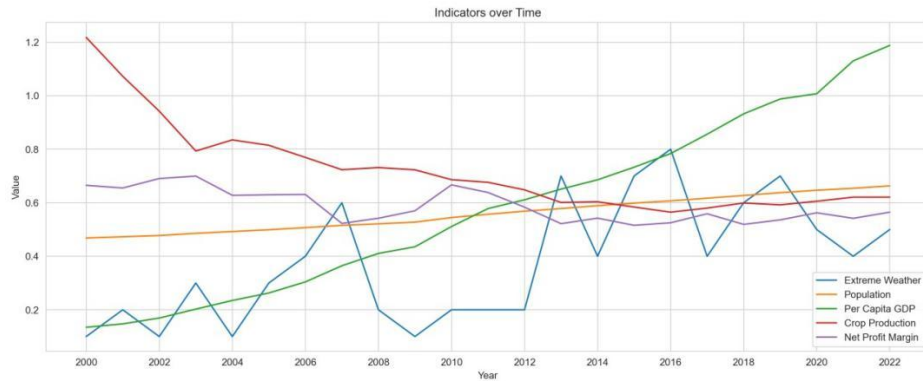

(a) Zhejiang

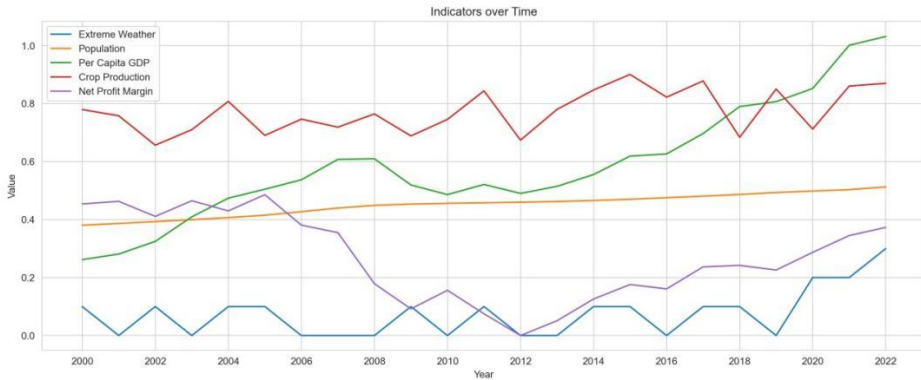

(b) Ireland

Figure 3: Trends of Various Indicators over Time

## 3.2 Model for Decision

### 3.2.1 Data Process

Due to significant differences in dimensions, magnitudes, and other aspects among different indicators, normalization of data is necessary. For computational simplicity, we have employed the straightforward yet effective min-max normalization method. This method ensures that all data is constrained within the range of [0,1]

$$x_{ij}^{t'} = \begin{cases} \dfrac{x_{ij}^t - x_{\min(j)}}{x_{\max(j)} - x_{\min(j)}}, & C_j \text{ is positive indicators} \\ \dfrac{x_{\max(j)} - x_{ij}^t}{x_{\max(j)} - x_{\min(j)}}, & C_j \text{ is negative indicators} \end{cases} \quad (2)$$

### 3.2.2 SMOTE Method

The SMOTE(Synthetic Minority Over-sampling Technique) is a data augmentation technique used to address class imbalance problems. It generates synthetic samples for the minority class by interpolating between the k nearest neighbors of each minority class sample. This helps balance the dataset and mitigates the bias towards the majority class in classification models. And the SMOTE algorithm is shown below:

---

**Algorithm 1 SMOTE Algorithm**

---

Require: Input data: Minority class samples $X_{major}$, majority class samples $X_{major}$, Number of nearest neighbors $k$, Number of synthetic samples to generate $N$
Ensure: Synthetic samples: $X_{synthetic}$

1: $X_{synthetic} \leftarrow \emptyset$

2: for $i \leftarrow 1$ to $N$ do

3:    $x = \text{Uniform}(X_{major})$    /*Select a random minority class sample*/ x

4:    $X_{top-k} = \text{Top-k}(X_{major}|\text{distance})$    /*Find $k$ nearest neighbors of $x$ in $X$*/

5:    $n = \text{Uniform}(X_{top-k})$    /*Select a random nearest neighbor $n$*/

6:    $x_{synthetic} = x + (n-x) \cdot \text{Rand}()$    /*Generate a synthetic sample $x_{synthetic}$ */

7:    $\text{Append}(x_{synthetic}, X_{synthetic})$    /*Add $x_{synthetic}$ to $X_{synthetic}$*/ return $X_{synthetic}$

---

To address the class imbalance problem, we applied the SMOTE algorithm to generate synthetic negative samples, ensuring a balanced dataset for SVM training. We used the SMOTE algorithm to artificially generate some negative samples. This way, we initially addressed the issue of imbalanced positive and negative samples in the data.

### 3.2.3 SVM

The basic idea of SVM learning is to find a separating hyperplane that can correctly classify the training dataset and maximize the geometric margin. For linearly separable datasets, there are infinitely many such hyperplanes (known as perceptrons), but the one with the maximum geometric margin is unique.Before deriving the solution, let's introduce some definitions. Suppose we have a training dataset on a feature space:

$$T = \{(x_1, y_1), (x_2, y_2), \ldots, (x_N, y_N)\} \tag{3}$$

Where $x_i \in R^n$, $y_i \in \{+1, -1\}, i = 1,2,\ldots N$, $x_i$ is the $i$th feature vector, and $y_i$ is the class label, with +1 indicating the positive class and -1 indicating the negative class. We assume that the training dataset is linearly separable.

The minimum value of the geometric margin $\gamma_i$ of the hyperplane with respect to all sample points is denoted as

$$\gamma = \min_{i=1,2\ldots,N} \gamma_i, \quad \gamma_i = y_i \left( \frac{w}{\|w\|} \cdot x_i + \frac{b}{\|w\|} \right) \tag{4}$$

In fact, this distance is the distance from the so-called support vectors to the hyperplane.Based on the above definitions, the problem of finding the maximum margin separating hyperplane in SVM can be formulated as the following constrained optimization problem:

$$\max_{w,b} \gamma \quad s.t. \quad y_i \left( \frac{w}{\|w\|} \cdot x_i + \frac{b}{\|w\|} \right) \geq \gamma, i = 1,2,\ldots, N \tag{5}$$

Dividing both sides of the constraint by $\gamma$, we obtain

$$y_i \left( \frac{w}{\|w\|\gamma} \cdot x_i + \frac{b}{\|w\|\gamma} \right) \geq 1 \tag{6}$$

Since $\|w\|$ and $\gamma$ are both scalars, for the sake of simplicity, we define

$$w = \frac{w}{\|w\|\gamma}, \quad b = \frac{b}{\|w\|\gamma} \tag{7}$$

$$\Rightarrow y_i(w \cdot x_i + b) \geq 1, \ i = 1,2,\ldots, N \tag{8}$$

Furthermore, since maximizing $\gamma$ is equivalent to maximizing $\frac{1}{\|w\|}$, which is equivalent to minimizing $\frac{1}{2}\|w\|^2$ (this is done to simplify the form for later differentiation, without affecting

the result), the problem of finding the maximum margin separating hyperplane in SVM can be formulated as the following constrained optimization problem:

$$\min_{\mathbf{w},b} \frac{1}{2} \|\mathbf{w}\|^2 \quad s.t. \quad y_i(\mathbf{w} \cdot \mathbf{x}_i + b) \geq 1, \ i = 1,2,\ldots,N \tag{9}$$

In practice, it is rare to encounter datasets that are completely linearly separable. To address this issue, the concept of "soft margin" is introduced, which allows for some points to not satisfy the constraints.

Using the hinge loss, the original optimization problem is reformulated as:

$$\min_{\mathbf{w},b,\xi_i} \frac{1}{2} \|\mathbf{w}\|^2 + C\sum_{i=1}^{m} \xi_i \quad s.t. \quad y_i(\mathbf{w} \cdot \mathbf{x}_i + b) \geq 1 - \xi_i (\xi_i \geq 0, i = 1,2,\ldots,N) \tag{10}$$

where $\xi_i$ is a "slack variable" defined as $\xi_i = \max(0, 1 - y_i(\mathbf{w} \cdot \mathbf{x}_i + b))$, which represents the degree to which a sample violates the constraints. $C > 0$ is the penalty parameter, and a larger value of C imposes a higher penalty on misclassifications.

By using the Lagrange multiplier method, we can obtain the Lagrangian function and subsequently derive its dual problem. Due to space limitations, we only provide the final objective function and constraints:

$$\min_{\mathbf{w},b,\xi_i} \frac{1}{2} \|\mathbf{w}\|^2 + C\sum_{i=1}^{m} \xi_i \quad s.t. \quad y_i(\mathbf{w} \cdot \mathbf{x}_i + b) \geq 1 - \xi_i (\xi_i \geq 0, i = 1,2,\ldots,N) \tag{11}$$

The optimal solution can be computed: $\boldsymbol{\alpha}^* = (\alpha_1^*, \alpha_2^*, \ldots, \alpha_N^*)^T$

Then we can obtain the optimal values for *w, b* and decision function for probability calculation:

$$\mathbf{w}^* = \sum_{i=1}^{N} \alpha_i^* y_i \mathbf{x}_i, \quad b^* = y_j - \sum_{i=1}^{N} \alpha_i^* y_i (\mathbf{x}_i \cdot \mathbf{x}_j) \tag{12}$$

## 3.3 Model for Prediction

### 3.3.1 C-D-C Function Based on C-D Model For Corresponding

The Cobb-Douglas production function, also known as C-D production function, is a commonly used economic model that describes the relationship between inputs and outputs in the production process.

The form of the C-D production function is as follows:

$$Y = \alpha K^{\beta_1} L^{\beta_2} \tag{13}$$

Taking the logarithm of the previous equation, we can obtain a linear equation:

$$lnY = \ln\alpha + \beta_1 lnK + \beta_2 lnL \tag{14}$$

In the formula, *Y* represents the growth rate of output; α represents the rate of technological progress; *K* represents the growth rate of capital; $\beta_1$ represents the capital output elasticity coefficient; *L* represents the growth rate of labor; $\beta_2$ represents the labor output elasticity coefficient. By generalizing the previous 2-dimensional formula and inspired by [3], we can obtain an n-dimensional formula:

$$lnY = \ln\alpha + \beta_1 lnK_1 + \beta_2 lnK_2 + \ldots + \beta_n lnK_n \tag{15}$$

In the formula, $Y$ is the dependent variable(Population or Per Capita GDP or Crop Production) $ln\alpha$ is an intercept constant, $K_j (j = 1,2\ldots n)$ represents various factors that affect $Y$, in our model they are the number of days of heatwaves, cold waves, heavy precipitation, drought, tornadoes each year, respectively. And $\beta_j (j = 1,2\ldots n)$ represents the coefficient of $lnK_j$.

By using Maximum Likelihood Estimation (MLE), we can estimate that when the weather index changes by 1%, other indicators, like $K_j$, will change by $\beta_j$.

3.3.2 Prediction and Decision

To evaluate the impact of extreme weather on decision-making, we can predict the values of other indicators by incorporating a certain change in the extreme weather indicator based on the previous year's indicators. Using the calculated values of $\beta_1$, $\beta_2$, and $\beta_3$ mentioned earlier, we can estimate the values of these indicators. the weather index changes by $\lambda\%$ :

$$x_{pred_j} = x_n \times (1 + \lambda \cdot \beta_j) \tag{16}$$

Finally, by incorporating these results into the decision model, we can obtain the final decision outcomes.

## 3.4 Implementation of The Model

3.4.1 Region Choose

In order to verify whether our model has a good evaluation and decision-making ability, we take Ireland in the European continent and Zhejiang Province, China in the Asian continent as examples to demonstrate our model.

- Ireland is located in the south-central part of the island of Ireland in western Europe, bordered by the Atlantic Ocean to the west and Northern Ireland to the northeast. Ireland's climate type is temperate maritime climate with abundant precipitation and it is often affected by Atlantic storms and experience extreme weather.

- Zhejiang Province, located in the southeast coastal area of China, has a monsoon humid climate. To the east of Zhejiang is a vast ocean, which provides landfall conditions for tropical cyclones in the Pacific Ocean. In addition, Zhejiang Province will also suffer from floods and droughts caused by meteorological anomalies.

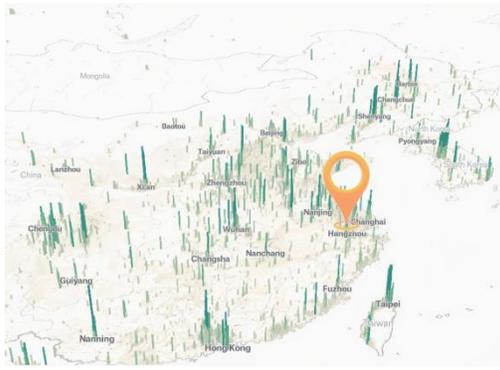 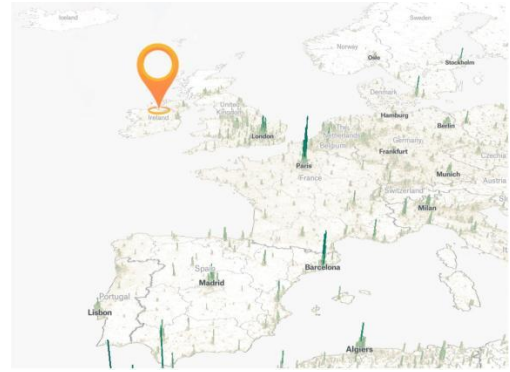

(a) Zhejiang　　　　　　　　　　　　　　(b) Ireland

Figure 4: Region Choose

### 3.4.2 Model Evaluation and Result Analysis

We collected data from 2000 to 2022, spanning a total of 23 years, as our sample. Among them, the year with the lowest net profit was considered as the negative sample. Using the SMOTE algorithm, we generated 23 synthetic data points, resulting in a total dataset of 46 samples. Due to the limited number of data points, we utilized k-fold cross-validation with $k = 5$ to calculate the average accuracy on the testing set as the final result. The average accuracy of the SVM model is 88.3% in Zhejiang and 79.6% in Ireland, demonstrating the model's effectiveness in predicting insurance policy probabilities under extreme weather conditions:

|  | Zhejiang | Ireland |
|---|---|---|
| Average Acc/% | 88.3 | 79.6 |

Table 2: Average Acc on Testing Set

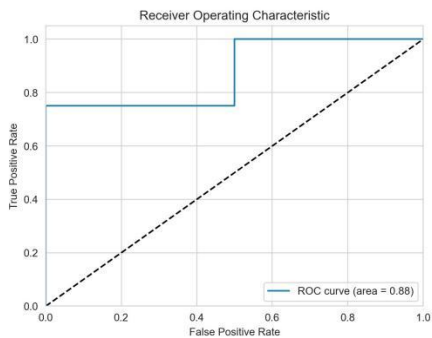 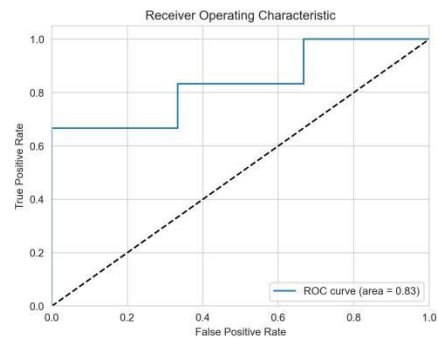

(a) Zhejiang　　　　　　　　　　　　　　(b) Ireland

Figure 5: One Fold's ROC Curve: Demonstrates that the classification performance is good

Now, we have utilized the C-D-C model to calculate the impact of extreme weather on other indicators, and the results are presented in the following table.

Subsequently, we computed the relationship between the change in extreme weather and the insurance probability. Specifically, for every *k*% variation in weather, the corresponding probability of insurance coverage is denoted as *p*.

|         | Crop Production/% | Per Capita GDP/% | Population/% |
|---------|-------------------|------------------|--------------|
| Zhejing | -0.9325           | -0.1891          | -0.2573      |
| Ireland | -0.4761           | -0.0042          | -0.0397      |

Table 3: The Impact of Extreme Weather on Other Indicators: Specifically, for every 1% change in extreme weather, there is a corresponding -0.9325% variation in crop yield.

Now, we can predict the change in insurance probability with the variation of extreme weather, as shown in the following figure:

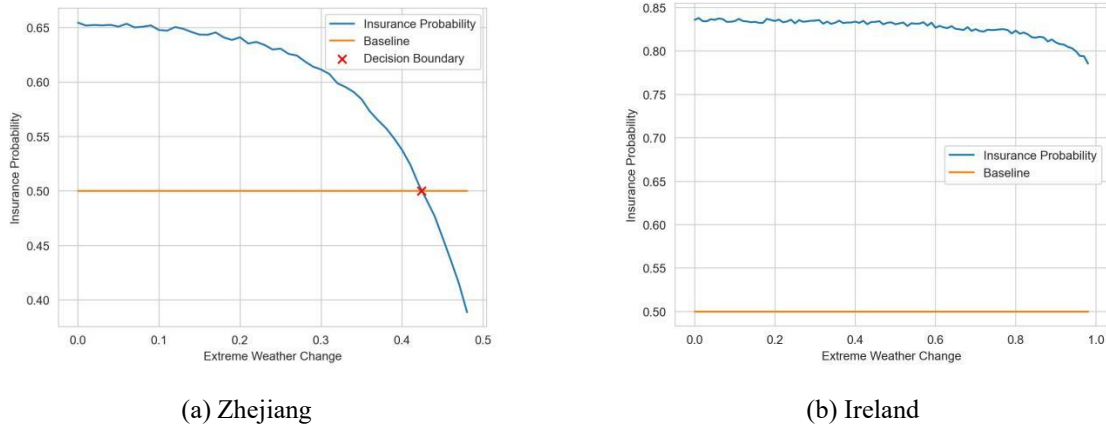

(a) Zhejiang  (b) Ireland

Figure 6: Trends of Underwriting Probability Indicators over Extreme Weather Changes

As the number of extreme weather events in Zhejiang increases, Fig.6(a) shows a gradual decrease in the predicted underwriting probability. When the number of extreme weather events reaches approximately 43%, the underwriting probability of the insurance company drops to 50%. This indicates a critical value for the underwriting probability in Zhejiang. It is believed that if the number of extreme weather events continues to rise, the insurance company will no longer provide underwriting services.

Fig. 6(b) illustrates the impact of extreme weather events. As the number of extreme weather events increases, the underwriting rate in the Irish region remains stable. However, even when the number of extreme weather events doubles, there is still a relatively high underwriting rate of about 78%. This suggests that extreme weather has a relatively minor impact on the underwriting issues of insurance companies in Ireland.

Further analysis was conducted on the main factors influencing the underwriting situation in the Irish region. Fig.6(b) demonstrates that both the per capita GDP of Ireland and the net premium margin experienced a significant decline in 2008, but returned to 2008 levels by 2016. Since then, both indicators have shown an upward trend. This analysis indicates a close relationship between the net premium margin of Ireland and the per capita GDP, suggesting that the Irish insurance industry is closely linked to the overall GDP. Therefore, our conclusion that extreme weather has a limited impact on the underwriting rate in Ireland is supported.

## 3.5  Possible Adventures

Based on previous research findings, it is not difficult to find that as the number of extreme weather events increases, the probability of insurance companies underwriting a certain area tends to decrease. However, this situation only considers the change in extreme weather and does not consider the impact of the joint change of extreme weather and other indicators on the underwriting probability. Tesselaar et al.'s research [7] has indicated that flood insurance drives population growth in European floodplains. Our data analysis reveals a positive correlation between the insurance net profit rate, population, and per capita GDP in Ireland since the global economic recovery in 2016. Despite considering multiple factors and our model's recommendation against insuring certain areas like Ireland, insurance companies may still take risks. This is because insuring certain policies can stimulate local population growth, leading to increased insurance net profit rates and greater profitability.

## 3.6  Property Owner

After fully understanding the regional insurance industry and real estate conditions, we believe that property owners can make some changes that affect the decision of whether or not an insurance company will choose to underwrite in that region:

- Property owners can choose to engage in the field of food production, promote the increase of food production, and attract more insurance companies to invest.

- Property owners can raise their awareness of insurance, reasonably increase the amount of insurance purchased, maintain the market share of insurance companies, and thereby maintain or even increase the investment ratio of insurance companies.

- Property owners can take preventive maintenance measures, such as scientifically reinforcing the house to improve its quality and thereby reducing the risk of the property, in order to increase its attractiveness to insurance companies.

# 4  House Construction: KSSC Model

We updated the model to address challenges faced by those without shelter due to population growth and extreme weather. The additions of indicators and K-Means made the model more specific to real estate investment. To contribute to the community, we expanded the insurance service area by adjusting population weight. Weather is no longer the sole factor, treated as a parallel indicator.

## 4.1  Region Choose

In order to better demonstrate our model and show that it has certain practical value, we choose to analyze the problem at the city level. During the analysis process, we fully considered the regional differences among cities, and ultimately selected a number of cities in Zhejiang Province, including (all cities of Zhejiang Province) as the analysis objects.

## 4.2 Upgrading Our Model: Adding Indicators and K-Means

### 4.2.1 Additional Factors Selection

We mainly added the following two indicators:

PCO is an essential criterion for assessing regional buildings stability. The higher the stability, the higher the degree of the regional houses resilience, the less impact natural disasters has on real estate investment and the greater the regional likelihood of insurability.

ARP is a important factor to measure the local population's demand for community and housing. The higher the ARP, the greater the demand for real estate developers and investors. If investors choose to insure in this area, they can balance both their interests and the needs of the community and population.

### 4.2.2 K-Means Algorithm

Unlike the previous model, we couldn't use the net profit margin to determine whether to build houses, there is no absolute relationship between the two. Therefore, we are using K-Means clustering to reassign positive and negative samples. This is understandable because locations suitable for building houses always have similar characteristics.

The K-Means algorithm can be briefly described in Algorithm 2:

---

**Algorithm 2 K-Means**

---

Require: Dataset $X = \{x_1, x_2, \ldots, x_N\}$, Number of clusters $K$

Ensure: $K$ Subset $X_k, k = 1, 2, \ldots, K$

 Randomly initialize $K$ cluster centers $C = \{c_1, c_2, \ldots, c_K\}$

1: while $\sum_{j}^{k} \frac{1}{|C_k|} \sum_{x_i \in C_k} ||x_i - c_k||_2 > \varepsilon$ do
2:  for $i \leftarrow 1$ to $N$ do
3:   $m = \mathrm{argmin}_j |x_i - c_j|^2$
4:   $x_i \rightarrow X_m$
5:  $c_k = \frac{1}{|C_k|} \sum_{x_i \in C_k} x_i$

 return $X_k$

---

### 4.2.3 Artificially Adjust the Weight of Population

In terms of subjective attitude, We want to offer appropriate services to the continuously growing population and community while ensure a certain level of profitability, we artificially increased the importance of population. Specifically, we amplified the population indicators of all cities by *k*% and input them into the decision-making model. Here we set *k* = 15 .

## 4.3 Analyze Zhejiang Province Using Our Model

### 4.3.1 Whether and Where to Build

For the cities in Zhejiang, the clustering results are shown in the following figure,(Using the original model in Task 1 as a benchmark ),Cluster 2 will be treated as positive examples, while Cluster 1 will be treated as negative examples. The subsequent processing will remain consistent with the previous model.

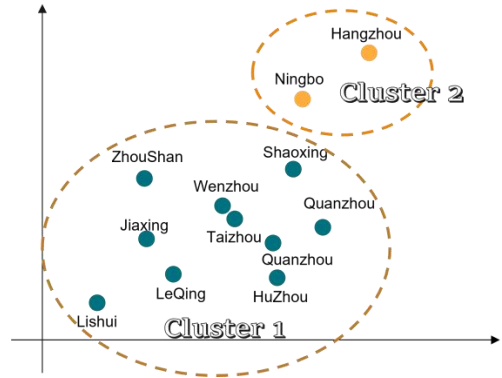

Figure 7: K-Means Results

About different investment regions, our upgraded model can cluster distinct regions to identify positive and negative examples, corresponding to constructing and non-constructing, respectively. Positive samples, before adjusting the population weight, represent cities with the highest potential for investment profit and should be invested in. After adjustment, if the samples change, regions with significant community demand are included as positive samples. To provide better service, they should also be considered for investment. The remaining cities have lower profit potential and medium community demand, and should not be invested in.

We adjust the weights of population. Surprisingly, the model's decisions underwent significant changes, as shown in the figure below. We find that Hangzhou, Ningbo, Shaoxing and Wenzhou are suitable cities for house construction.

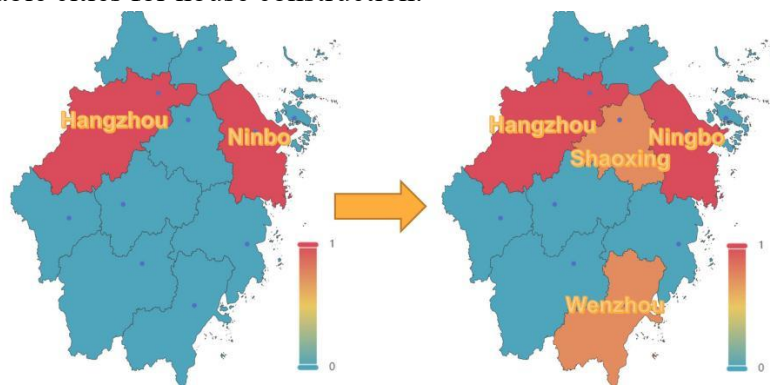

Figure 8: Where to Build

4.3.2 How to Build

We have calculated the mean of each indicator data:

$$\bar{x}_j = \frac{\sum_{i=1}^{m} x_{ij}}{m} \tag{17}$$

And the deviation of the score of each indicator from the average score of that indicator:

$$\Delta x_{ij} = x_{ij} - \bar{x}_j \tag{18}$$

After analyzing the shared characteristics of positive and negative city samples, we observed that the deviation of extreme weather values from the average is lower than zero in the positive sample areas, and the corresponding value in PCO is higher than zero. Consequently, we can deduce the following recommendations:

- high-quality construction: Emphasize the selection of building materials and construction processes to ensure reliable, durable and eco-friendly properties to improve PCO.
- Sustainable development: Focus on environmental sustainability, adopt energy-saving and emission-reduction measures such as green buildings, striving to minimize the impact of human-induced extreme weather
- Community facilities: Provide various community facilities and services such as green spaces, sports facilities, children's playgrounds, etc. to enhance the attractiveness and cohesion of the community and better service to people.

## 5 Preservation Model: TOA Model

In this Section, we have developed the TOPSIS -ORM and AHP model to evaluate and assign scores to various types of landmarks in the community. We have selected a set of ordinary buildings as reference points and calculated their average scores. If the score of a historical building is higher than this average score, it is considered worthy of preservation; otherwise, it is deemed unnecessary to preserve. Based on the importance of different criteria, we have formulated policies to protect the buildings within the community and quantitatively described the impact of these policies.

### 5.1 Landmark and Indicators Choose

We selected 16 landmarks in 14 regions of China and five major indicators about these buildings. And choose five important indicators to evaluate the value of landmarks: Existing time ,Floor space ,Annual tourist reception, Number of documents, Cultural value score(Use GPT-4 to evaluate).

The landmarks we select are shown in the Fig.9.

## 5.2 Model for Score: TOA Model

### 5.2.1 TOPSIS-ORM

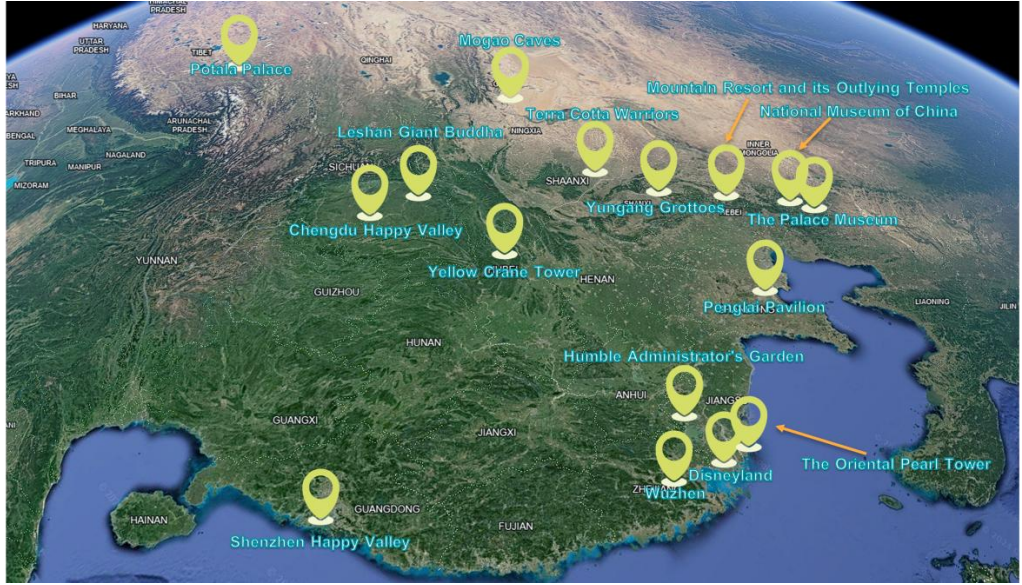

Figure 9: Landmarks Choose

Step 1. Calculation of Indicator Importance By TOPSIS

For a given indicator, if it interplays greatly between indicators within an index system, then it can be considered as an important factor for the special evaluation object. In this case, relatively greater weight should be assigned to this indicator. To calculate the interaction, we use the reciprocal Hamming distance, the larger the $D_{ij}$, the greater the interaction:

$$D_{ij} = \frac{m}{\sum_{k=1,k\neq j}^{m-1} |x_{ik}' - x_{ij}'|} \tag{19}$$

We will now determine the importance of each indicator. Calculate the best and worst solution of interaction for indicator $j$, then calculate the Euclidean distance between $D_{ij}$ and best or worst solution:

$$\begin{cases} D_j^+ = \max_i\{D_{1j}, D_{2j}\cdots D_{nj}\} \\ D_j^- = \min_i\{D_{1j}, D_{2j}\cdots D_{nj}\} \end{cases} \Rightarrow \begin{cases} D_j^{ideal} = \sqrt{\sum_{i=1}^n (D_{ij}^t - D_{ij}^+)^2} \\ D_j^{poor} = \sqrt{\sum_{i=1}^n (D_{ij}^t - D_{ij}^-)^2} \end{cases} \tag{20}$$

$$(i = 1,2,\cdots n; j = 1,2,\cdots, m)$$

Using the results above, we can determine the importance of each indicator $S_j$:

$$S_j = \frac{D_j^{poor}}{D_j^{ideal}+D_j^{poor}} \tag{21}$$

Obviously, $S_j \in [0,1]$, therefore, when $S_j$ tends to 1, it indicates that the indicator is more important for landmark.

Step 2. Calculation of Indicator Weights By ORM

Order Relation Method (ORM) is a method used for multi-criteria decision-making, focusing on establishing order relationships among different criteria. The goal of this method is to determine the relative superiority or inferiority among different criteria or objects, thereby supporting the final decision-making process. We use ORM to calculate indicators' weights:

First Determine the descending ratio of indicator importance

$$\frac{S_{j-1}}{S_j} = r_j, j = m, m-1, \cdots, 3, 2 \tag{22}$$

Then calculate the indicator weights $\Sigma w_j = 1$, and the proof can be found in [2]):

$$w_m = (1 + \sum_{j=2}^{m} \prod_{i=j}^{m} r_i)^{-1}, w_{j-1} = r_j w_j, j = m, m-1, \cdots, 3, 2 \tag{23}$$

5.2.2 AHP Method

Analytic Hierarchy Process (AHP) is an accurate approach for quantifying the weights of decision criteria, widely used in different decision situations, in fields like government, business, healthcare, shipbuilding, industry and education. In our analysis, we use this method to determine weights of different main factors. Similarly, we determine the weights of different indicators in each factor. The following are the steps.

Step 1. Factors Selection

We have selected several main factors previously. And you can find them in 5.1

Step 2. Calculation of Comparison Matrix

Due to a lack of relevant knowledge, we leveraged GPT-4 to construct our comparison matrix $(b_{ij})_{5\times5}$, where the rows and columns respectively represent Existing time, Floor space, Annual tourist reception, Number of documents, Cultural value score, and $F_{ij}$ represents the degree of importance of $F_i$ relative to $F_j$.

$$\begin{bmatrix} 1 & 3 & 2 & 4 & 1 \\ \frac{1}{3} & 1 & \frac{1}{2} & 2 & \frac{1}{3} \\ \frac{1}{2} & 2 & 1 & 3 & \frac{1}{4} \\ \frac{1}{4} & \frac{1}{2} & \frac{1}{3} & 1 & \frac{1}{2} \\ 1 & 3 & 4 & 2 & 1 \end{bmatrix} \qquad (24)$$

Step 3. Consistency Test

We can calculate the eigenvalues and eigenvectors of the matrix before. Next we need to perform consistency test with the maximum eigenvalue:

$$CI = \frac{\lambda_{max} - n}{n-1}, CR = \frac{CI}{RI} \qquad (25)$$

where $RI$ = 1.12 when $n$= 5. For the above comparison matrix, we obtain $CR$= 0.06 ≤ 0.1, thus the comparison matrix is acceptable.

Step 4. Calculation of Weights

Passing the consistency test, we can get the weights of main factors by the eigenvector corresponding to the maximum eigenvalue: Existing time(0.3383) ,Floor space(0.1078) ,Annual tourist reception(0.0826),Number of documents(0.1636),Cultural value score(0.3383)

5.2.3 Combined Weights and Scoring

We use the linear superposition method to combine weights, resulting in the final weights for each indicator, Where α is hyper parameters, and in this case, we set α= 0.5.:

$$z_j = \alpha w_j + (1 - \alpha) a_j \qquad (26)$$

Then we can assign scores to each city:

$$Scores_j = x_j \cdot z^T \qquad (27)$$

## 5.3 Model Implementation

5.3.1 Weights for Indicators

Weights emphasize the contribution and importance of various factors or indicators. By using the model we established to analyze the weights among indicators, we obtained the weights of each indicator. Among them, the indicator with the highest weight is Cultural value score, while the one with the lowest weight is Annual tourist reception. And the following figure demonstrates all the weights:

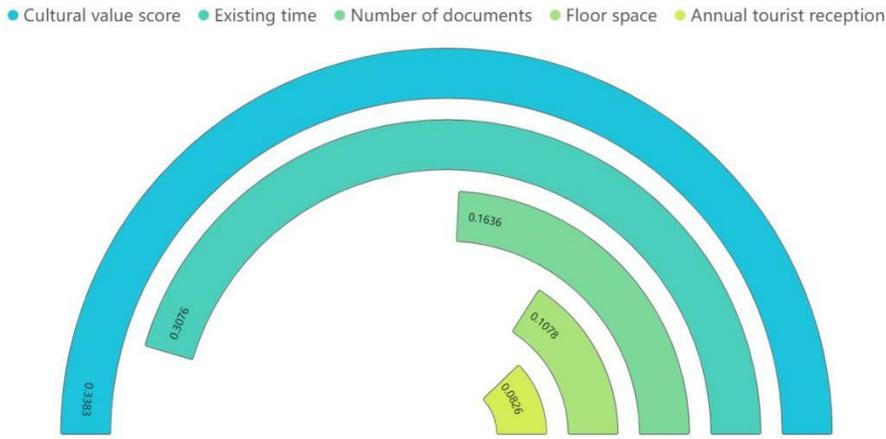

Figure 10: Weights for Indicators

### 5.3.2 Scores for Buildings

The scores of architectural properties determine the ranking of protective importance. The Palace Museum ranks first with a score of 0.93, indicating the highest level of protection needs and value. Chengdu Happy Valley and Shenzhen Happy Valley have the lowest scores, with 0.28 and 0.29 respectively. The 16 buildings can be categorized into three gradients based on their scores: the first gradient (ranked one to five) with scores above 0.7, the second gradient (ranked six to twelve) with scores between 0.5 and 0.7, and the third gradient (ranked thirteen to sixteen) with scores below 0.5.

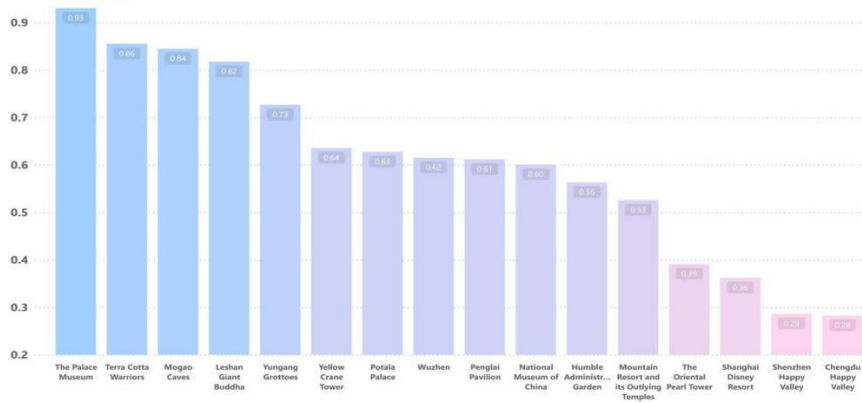

Figure 11: Scores for Landmarks

### 5.3.3 Conservation Value Determination

For different architecture, using our model to identify its *score*. If *score* ≥ 0.5, it needs to be protected; If *score* < 0.5, then there is no need.

### 5.3.4 Measure Should Be Taken

- Simply because of the significant weight of history and culture, it is the top priority to promote the cultural history of the building and increase cultural identity.

- And due to ATR and floor space also matter, improving the tourism industry in the area surrounding the building to increase its economic value and floor space is advised.

- What's more, the restoration and protection of the building itself are essential undoubtedly.

# 6 Nanxun Ancient Town: Protect It!

## 6.1 Landmark Choose

Huzhou City is located in the northern part of Zhejiang Province and is prone to the impact of extreme weather such as heavy rainfall and typhoons. Therefore, after fully considering our model and the available data, we selected the ancient town of Nanxun in Huzhou City as the research object for a detailed analysis.

## 6.2 Model Implementation

The results and analysis of the Nanxun Town in the insurance and protection models are as follows:

- In SSC-Insurance Model, the probability of the ancient town of Nanxun being insured is 65.32%>50%, which Means that the insurance company will insure this building.

- In TOA-Preservation Model, the score of the ancient town of Nanxun is 0.512, which Means that the ancient town of Nanxun has certain protection value. However, the score is close to the threshold value of the protection model, and during the fluctuation of indicators, it may cause the score to fall below the decision threshold of 0.5, resulting in the building not being protected.

## 6.3 Brief Insight and Measures

- Short-term Measures: Restore and protect ancient towns.
- Medium-term Measures: Increase publicity efforts to enhance cultural influence.
- Long-term Measures: Enhance the historical and cultural depth of ancient towns

# 7 Robustness Analysis

Due to the subjective selection of weighting factors in AHP and TOPSIS-ORM, as well as the limited information on certain landmarks in the training dataset of GPT-4, the scoring of cultural values for landmarks may also have some biases, which could affect the accuracy of the model. Therefore, we added a normally distributed random error $\varepsilon$, $\varepsilon \sim N(0,0.5)$ to the normalized data. We then compared the scores obtained with the original results to examine the robustness of our model. And the results are shown below, The results shown in the graph indicate that our model exhibits good robustness.:

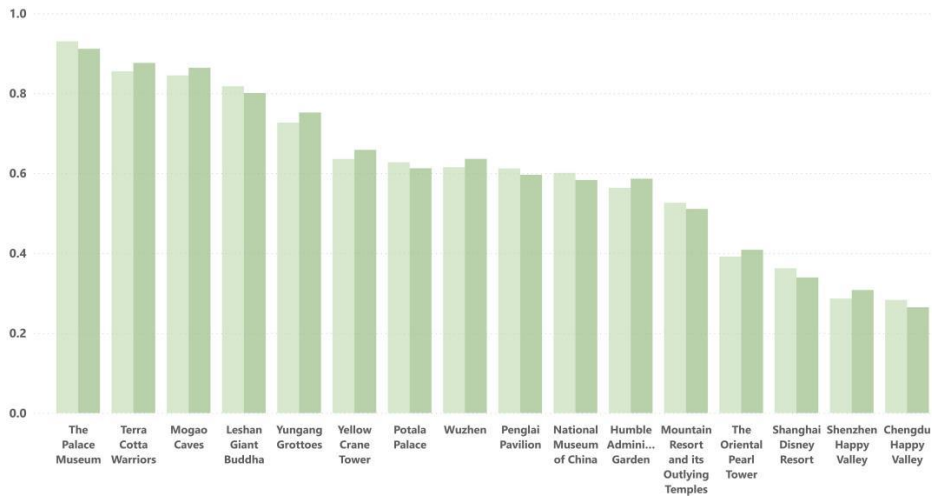

Figure 12: Comparison of Scores Before and After

# 8 Model Strengths and Weaknesses

## 8.1 Strengths

- The advantages of the SMOTE addresses class imbalance problems by synthesizing minority class samples, increases the number of minority class samples to improve the classifier's ability to recognize them.
- TOA-Preservation Model is proved to have strong robustness, which verifies the accuracy of our model.
- By applying the C-D-C method, we found the relationship between extreme weather and population, per capita GDP and food production, making our model more reasonable and accurate.

## 8.2 Weaknesses

- In the process of using SVM for modeling, we subjectively believe that this is a biased linear problem and can obtain more ideal results by using SVM. In fact, this problem may be a more complex nonlinear problem, and the treatment of nonlinear problems by SVM may deviate from expectations.
- Our information channels are limited, and even though we later expanded the limited sample size, we still believe that with a larger sample set, the model performance would be better.

# 9 Conclusion

Extreme weather escalation intensifies conflict between insurers' gains and homeowners' financial burdens, so we establish a SSC-Insurance Model, integrating SMOTE, SVM, and C-D-C, to enhance insurance sector durability and safeguard real estate interests. TOA-Preservation Model is also built, customizing solutions by evaluating buildings' protection requirements in affected regions. Here is what we do:

- Introducing the SSC-Insurance Model, we achieved 88.3% accuracy in Zhejiang and 79.6% in Ireland using SVM. A 43% increase in extreme weather events serves as the threshold for insurance purchase in Zhejiang, while weather has minimal impact on insurance decisions in Ireland.

- By incorporating 2 indicators and employing K-Means, our model becomes more specific to real estate investment. We expand the insurance service area by adjusting the population weight, identifying Hangzhou, Ningbo, Shaoxing, and Wenzhou as suitable cities for real estate development in Zhejiang. We encourage developers to built high-quality, sustainable, and community-oriented projects.

- Developing a protection model utilizing TOPSIS-ORM and AHP, we assess 16 buildings in 14 Chinese regions. Cultural value carries the highest weight(0.3383),while tourist visits have the least significance(0.0826).Buildings are divided into three categories for protection.

- Nanxun Ancient Town in Zhejiang Province serves as our focal point. Our insurance model predicts a 65.32% probability of coverage, while our protection model assigns it a score of 0.512.

# 10 Letter to Nanxun Community

Dear leaders of Nanxun Community,

Nanxun Ancient Town, located in Nanxun District, Huzhou City, Zhejiang Province, is a town with abundant cultural resources and a blend of Chinese and Western architecture, carrying significant cultural, historical, economic, and community values. However, Zhejiang Province is an area prone to extreme weather such as typhoons and floods, which may have an impact on the protection of Nanxun Town, for instance, insurance coverage issues. Therefore, our team established SSC-Insurance Model to analyze and predict the related risks of insurance coverage. At the same time, we established a TOA-Protection Model on protection issues to analyze various indicators that affect the protection value of the building.

We proposed the SSC-Insurance model, which assesses underwriting risks and aids in decisionmaking. Using the collected data, we enlarged the sample set using SMOTE and trained SVM to predict changes in insurance coverage probability with increasing extreme weather events. Our results indicate that insurance coverage probability remains strong for Nanxun Ancient Town unless extreme weather events exceed 43%.

We then proceeded to analyze the protection value of Nanxun Ancient Town. Based on TOPSISORM and AHP, we established TOA-Preservation Model that can evaluate the protection value of buildings. We collected five indicators for Nanxun Ancient Town: history, occupied area, tourist reception volume, number of documents, and cultural value score. The final score for Nanxun Ancient Town was 0.512, slightly above the decision threshold of 0.5. To enhance the protection value of Nanxun Ancient Town and conserve our material and cultural heritage, our team offers the following recommendations:

- In the short term, it is necessary to comprehensively restore the old houses and buildings within the ancient town, effectively restore important historical buildings with cultural value, and maintain the authenticity of history and the integrity of the ancient cityscape while continuing to fulfill the functions of the ancient town.

- In the medium term, it is important to strengthen promotional efforts for the ancient town, focus on developing tourism-related industries around Nanxun Ancient Town to drive economic development in the surrounding areas, and fully leverage regional synergies.

- In the long term, it is essential to fully tap the historical and cultural resources of the ancient town, tell cultural stories, protect the region's non-material cultural heritage, and comprehensively enhance the cultural depth and historical thickness of Nanxun Ancient Town.

- In terms of cost proposals, it is recommended to fully utilize available funds for rational planning and investment, while also purchasing insurance reasonably to address possible emergencies.

Sincerely,

L Qin, D Hua andY Xie

# A  Report AI USE

1. OpenAI GPT-4 (Mar 14, 2023 version, GPT-4)

Query: Assuming you are now a community leader, tasked with determining which buildings in the community should be protected and preserved due to their cultural, historical, economic, or community significance. Let's consider five criteria: building lifespan, building footprint, number of literature references in the database, visitor traffic, and historical-cultural significance. Please find below the comparison matrix for these five criteria use AHP, return the matrix in LaTeX format

Output: As a community leader, I will provide a comparison matrix using the Analytic Hierarchy Process for the five indicators - building age, building size, literature documentation, visitor traffic, and historical-cultural significance - to determine the buildings that should be preserved and protected in the community.

|  | Building Lifespan | Building Footprint | Literature References | Visitor Traffic | Historical-Cultural Significance |
|---|---|---|---|---|---|
| Building Lifespan | 1 | 3 | 4 | 3 | 5 |
| Building Footprint | $\frac{1}{3}$ | 1 | 2 | 2 | 3 |
| Literature References | $\frac{1}{4}$ | $\frac{1}{2}$ | 1 | 3 | 4 |
| Visitor Traffic | $\frac{1}{3}$ | $\frac{1}{2}$ | $\frac{1}{3}$ | 1 | 3 |
| Historical-Cultural Significance | $\frac{1}{5}$ | $\frac{1}{3}$ | $\frac{1}{4}$ | $\frac{1}{3}$ | 1 |

The above is a comparison matrix, where 1 represents equal importance between two indicators, and the numerical ratios represent the relative importance of one indicator to another. Please note that this is just an example, and the actual comparison matrix may need to be adjusted based on specific circumstances.

2. OpenAI GPT-4 (Mar 14, 2023 version, GPT-4)

Query:Now you are a professional historian, and I need you to rate some attractions from a cultural value perspective, ranging from 1 to 10. Please ensure that the ratings have distinct differences. You only need to provide the ratings, no additional output is required. The attractions that need your ratings are: Mogao Grottoes, Forbidden City, Terracotta Army of Qin Shi Huang, Potala Palace, Humble Administrator's Garden, Chengde Mountain Resort, Yungang Grottoes, National Museum of China, Yellow Crane Tower, Penglai Pavilion, Wuzhen Water Town, Leshan Giant Buddha, Oriental Pearl Tower, Shanghai Disneyland, Happy Valley in Shenzhen, Happy Valley in Chengdu, and Nanxun Ancient Town.

Output:

Mogao Caves:10

The Palace Museum:10

Terra Cotta Warriors:10

Potala Palace:8

Humble Administrator's Garden:8

Mountain Resort and its Outlying Temples:7

Yungang Grottoes:10

National Museum of China:9

Yellow Crane Tower:6

Penglai Pavilion:8

Wuzhen:6

Leshan Giant Buddha:9

The Oriental Pearl Tower:4

Shanghai Disney Resort:3

Shenzhen Happy Valley:3

Chengdu Happy Valley:3

Nanxun Ancient Town:7